\documentclass[preprint,amsmath,amssymb,aps,prd,showpacs]{revtex4-1}

\usepackage{epsfig}
\usepackage{dcolumn}
\usepackage{bm}
\usepackage{tikz}
\usepackage{graphicx, graphics, color}
\usepackage{dcolumn}
\usepackage{bm}
\usepackage{hyperref}
\usepackage[mathlines]{lineno}

\newcommand{\be}{\begin{equation}}
\newcommand{\ee}{\end{equation}}
\newcommand{\ben}{\begin{eqnarray}}
\newcommand{\een}{\end{eqnarray}}

\newcommand{\cH}{{\cal H}}
\newcommand{\cE}{{\cal E}}

\newcommand{\p}{\partial}
\newcommand{\na}{\nabla}

\newcommand{\te}{\tilde e}

\newcommand{\tF}{\tilde F}

\newcommand{\tg}{\tilde g}

\newcommand{\ep}{\epsilon}

\newcommand{\ga}{\gamma}

\newcommand{\tB}{{\tilde B}}
\newcommand{\tA}{\tilde A}
\newcommand{\tE}{{\tilde E}}

\begin{document}

\title{Black hole dark monopole system}
\author{Marek Rogatko} 
\email{rogat@kft.umcs.lublin.pl}
\affiliation{Institute of Physics, 
Maria Curie-Sklodowska University, 
20-031 Lublin, pl.~Marii Curie-Sklodowskiej 1, Poland}

\date{\today}

\begin{abstract}
We scrutinize properties of electrical charges bounded to visible and dark matter sectors, in the vicinity of a magnetic poles of both sectors.
It turns out that the considered system 
has an angular momentum despite the charges are at rest.
On the other hand, investigation the behavior of electric charges of visible and hidden sectors
 held at rest outside a magnetically charged black hole, discloses
that even if the electric charge are regarded as perturbations on a spherically symmetric magnetic static black hole in  dark photon theory,
at large distances it looks like a stationary axisymmetric magnetically charged black hole.
\end{abstract}

\maketitle

\section{Introduction}
For a long period the influence of electromagnetic field on black hole spacetime has been attracted much attention to.
Namely, the point charge in the vicinity of Schwarzschild black hole was considered in \cite{han73}-\cite{coh71}, while
the electrostatic and magnetostatic in the spacetimes of Schwarzschild and Reissner-Nordstr\"om, were described in \cite{lin76}.
On the other hand, magnetic current loops in the vicinity of static black hole were analyzed in \cite{pet74}.

The problem of a point charge at rest near a Reissner-Nordstr\"om black hole has been elaborated in Refs. \cite{gerruf}, while 
the external stationary magnetic fields in the presence of Kerr black holes using the test field approximation in Ref. \cite{wal74}.

In \cite{bic80} it has been revealed that 
as electromagnetic multipole moments approach the event horizon of a black hole, all the multipole moments fade away except the monopole one.
An interesting result concerning the component of the electromagnetic field, which claimed that all normal to the event horizon components disappeared as one considers 
extremal black hole and flux lines were totally expelled from such black hole, was found in Ref. \cite{cha98}.

In the last few years, there have been vast amount of attempts to study the motion of particles around various classes of black holes.
If one restricts to the case of static black hole examples, there have been considered such topics as spinning particles motion in background of Schwarzschild  black holes \cite{abd23},
charged particles and their oscillations around magnetized Schwarzschild black hole \cite{qi23},  motion and collisions of particles with magnetic dipole moment and electric charge in dipolar magnetosphere around static black holes \cite{ull24}. On the other hand,
the problem of both charged \cite{fro10}-\cite{,zah13} and magnetized \cite{fel03} particle motion in the weakly magnetized Schwarzschild,
non-Schwarzschild \cite{ray16}, and Gauss-Bonnet  \cite{abd20} black holes spacetime was revealed.

As far as the {\it dark sector} is concerned, magnetized particle motion and collision process analysis of two particles 
in the spacetime of a static black hole influenced by weak magnetic fields stemming from visible and 
{\it dark matter} sectors, have been analyzed in \cite{rog25a}. 
The synchrotron radiation of a massive charged under visible and hidden sector groups, moving in equatorial plane around spherically symmetric weakly magnetized black hole
was studied in \cite{rog25b}. It was found that the radiation power and energy loss of the particle might constitute the possible imprint of dark matter.

In addition, an enthralling feature of an electrical charge in the vicinity of a magnetic pole has been found. To be specific,
in Ref. \cite{gar91} it was revealed that the total angular momentum dependent on the separation distance between magnetically black hole an electric charge
might give rise to the possibility to spin up the black hole in question, when it moved closer to its event horizon.
It was also remarked that in the case of an extremal Reissner-Norstr\"om magnetic black hole the spinning process was not possible until the the charge was outside the event horizon.

The more complicated models of black holes were also taken into account. Namely,
in \cite{kim07}, the angular momentum of an electric charge in the vicinity of magnetically charged dilaton black hole, being the solution of the low-energy string theory,
was elaborated. As in \cite{gar91} in the extremal limit of the studied black hole, the angular momentum tends to be independent of the distance black hole-electric charge.

The problem of an electric charge magnetically charged black hole system, has been considered in \cite{bun07}, where the ADM formalism is implemented to the studies.
Moreover, the Taub-NUT spacetime in the context of electric charge plus magnetic black hole was elaborated. The main conclusions were the same as in the previous works \cite{gar91,kim07}.

On the other hand, the problem of magnetic monopoles in {\it dark photon} theory, in the context of minicharged particles which arise naturally in the extension of SM has been elaborated
in \cite{bru09}.

In our paper we shall focus on the problem how physics beyond the Standard Model (SM), including {\it dark matter}, will modify the possible spinning up black hole process.
To be more precise we shall use the {\it dark photon} model, which action is provided by
\be
S_{M-dark~ photon} = \int  \sqrt{-g} d^4x  \Big(
- F_{\mu \nu} F^{\mu \nu} - B_{\mu \nu} B^{\mu \nu} - {\alpha}F_{\mu \nu} B^{\mu \nu}
\Big),
\label{ac dm}
\ee  
where $\alpha$ denotes
the {\it kinetic mixing} parameter connected with the interaction strength between {\it dark photon} and Maxwell one.
Recently
the new limit for it, $\alpha =1.6 \times
10^{-9}$ and the mass range of {\it dark photon}
$2.1\times 10^{-7} - 5.7 \times10^{-6} eV$, have been found in Ref. \cite{fil23}. Moreover,
quantum limited amplification enable for the first time, to establish the kinetic mixing coupling constant to $10^{-12}$ level for majority of {\it dark photon} masses \cite{ram23}. 
An upper bound on 
kinetic mixing parameter
 $\alpha < 0.3-2 \times 10^{-10}$ (at 95 percent confidence level) has been found in \cite{kot23}.

The model represent the simple extension of the SM, introducing the auxiliary
$U(1)$ Abelian gauge boson which couples to the ordinary Maxwell gauge field via {\it kinetic mixing term} \cite{hol86,cap21}.

The justification for this model stems from
the contemporary unification scheme  \cite{ach16}, where there exist mixing portals coupling the ordinary Maxwell 
and auxiliary gauge fields charged under their own groups. In addition several high-energy non-typical astrophysical observations like for instance like $511~ keV$
gamma rays \cite{jea03}, the excess of the positron cosmic ray flux in galaxies \cite{cha08} or the observations of an anomalous monochromatic 
$3.56~ keV$ X-ray line in the spectrum of some galaxy clusters \cite{bub14}, can be explained in the realm of the model in question.
 
Nowadays,  there can be also observed the ongoing search using astrophysical and laboratory experiments \cite{fil20}, in order
to find the range of values for {\it dark photon} -Maxwell field  $\alpha$-coupling constant, as well as, the range for the possible mass of the {\it hidden photon}. 
Gamma rays emission studies from dwarf galaxies \cite{ger15},
inspection of dilaton-like coupling to photons caused by ultra-light {\it dark matter} \cite{bod15}, scrutinising the
 fine structure constant oscillations \cite{til15}, {\it dark photon} emission during supernova event \cite{cha17}, 
 electron excitation measurements in CCD-like detector \cite{sensei}, the search for a {\it dark photon} in $e^+ e^-$ collisions at BABAR experiment \cite{lee14},
 measurements of the muon anomalous effect \cite{dav11}, are the only few examples of the searches in progress.
 Recently the name {\it dark astronomy} was proposed to 
examine the detection prospects of a dark radiation signal originating from a dissipative dark matter sector, i.e., a massive {\it dark photon}
was assumed to mediate the dissipative interactions among {\it dark matter} particles \cite{alo25}. The dark radiation flux could be presumably detectable for
underground experiments XENON/SENSEI.

On the other hand, the lack
 of  evidence of the most popular {\it dark matter} candidates in the experimental data, 
 precipitates to diverse searching methods
to elucidate {\it dark sector} \cite{ber18b}. 
It turns out that
one of the most promising way of finding the presence of {\it dark matter} are environmental studies of compact objects like black holes,
 wormholes and compact star-like objects. Some work in this direction, 
 connecting with {\it dark matter} clouds and their influence on compact objects were performed in 
 \cite{kic19}-\cite{kic22}. 
In view of the aforementioned problems, the main reason of our studies will be addressed to the environment investigations of {\it dark photon} theory
in the vicinity of magnetically charged black hole.

The organization of the paper is as follows. In Sec. II we propose the transformation which 
enables one to get rid of $\alpha$-coupling constant from the equations of motion, by the redefinition of the $U(1)$-gauge field components. In Sec. III
we elaborate the problem of {\it dark photon} monopole and obtain the generalization of Dirac quantization conditions for the case of {\it dark sector}.
Sec. IV is connected with the possible observational consequences of {\it dark monopole} existence, while
Sec. V will be devoted to the existence problem of electrical charges, pertaining to both $U(1)$-gauge groups, in the nearby of 
magnetically charged static spherically symmetric black hole. 

In Sec. VI we discuss the observational difference in rotation parameter of stationary axisymmetric black hole, caused by {\it dark sector} monopole.
Sec. VII concludes our investigations.
In our paper we use geometrized units $G=1,~c=1$, while the line element has the signature $(- +++)$.

\section{Dark photon coupling} 
\label{sec:darkphoton}
In this section we propose a transformation of underlying gauge fields which enables one to get rid of {\it kinetic mixing} term. 
Its application results in obtaining new gauge fields, which are mixture of $A_\mu$ and $B_\mu$ ones, in addition with the adequate factors containing
$\alpha$-coupling constant.
In the next step we proceed to derive the adequate equations of motion for the underlying theory.

In order to eliminate the {\it kinetic mixing } term, 
from the action (\ref{ac dm}),
 we define new gauge fields, being the mixture of the initial ones appearing in the action (\ref{ac dm})
\ben \label{transA}
\tA_\mu &=& \frac{\sqrt{2 -\alpha}}{2} \Big( A_\mu - B_\mu \Big),\\ \label{transB}
\tB_\mu &=& \frac{\sqrt{2 + \alpha}}{2} \Big( A_\mu + B_\mu \Big).
\een
By virtue of the above relations one arrives at the following:
\be
 F_{\mu \nu} F^{\mu \nu} +
B_{\mu \nu} B^{\mu \nu} +  \alpha F_{\mu \nu} B^{\mu \nu}
\Longrightarrow
 \tF_{\mu \nu} \tF^{\mu \nu} +
\tB_{\mu \nu} \tB^{\mu \nu},
\ee
where one denotes  $\tF_{\mu \nu} = 2 \na_{[\mu }\tA_{\nu ]}$, and respectively $\tB_{\mu \nu} = 2 \na_{[\mu }\tB_{\nu ]}$. Consequently the action (\ref{ac dm})
can be rewritten in the form as
\be
S_{M-dark~ photon}  = \int \sqrt{-g} d^4x  \Big(
- \tF_{\mu \nu} \tF^{\mu \nu} - \tB_{\mu \nu} \tB^{\mu \nu}
\Big).
\label{vdc}
\ee
Variation of the action (\ref{vdc}) with respect to $g_{\mu\nu},~\tA_\mu$ and $\tB_\mu$ reveals the equations of motion for Maxwell-{\it dark matter} system. Namely they are given by
\be
\na_{\mu} \tF^{\mu \nu } = 0, \qquad \na_{\mu} \tB^{\mu \nu } = 0.
\label{fb}
\ee
One can remark that such redefinition of the gauge fields (like in equations (\ref{transA})-(\ref{transB}) can be always done in the case of $U(1)$-gauge groups.

To proceed further,
having in mind relations (\ref{vdc}) and (\ref{fb}), for the consistency of the presented approach one requires the appropriate redefinitions of the charges coupled to original fields.
Accordingly they are defined as follows:
\ben \label{cA}
\te&=& \frac{\sqrt{2 -\alpha}}{2} \Big( e - e_d \Big),\\ \label{cB}
\te_d &=& \frac{\sqrt{2 + \alpha}}{2} \Big( e + e_d \Big).
\een
In the above equations (\ref{cA})-(\ref{cB}), $e$ stands for the Maxwell charge, while $e_d$ is connected with {\it dark sector} one. 

Due to the aforementioned transformations of charges and gauge fields we have a mixture of $A_\mu$ and $B_\mu$ fields, appearing in the relations (\ref{transA})-(\ref{transB}),
as well as, combinations of electric Maxwell and {\it dark} electric charges. Although $\alpha$-constant disappear in the equation (\ref{vdc}), by the redefinitions of the fields,
it is still present in the definitions given by (\ref{cA})-(\ref{cB}) and  (\ref{transA})-(\ref{transB}).

\section{Dark photon Monopole angular momentum}
The recent investigations envisage that the
different candidates for the theory which generalizes the Standard Model usually assumes the several stages
of symmetry breaking bounded with the emergence of {\it magnetic monopoles} \cite{dark mon}.

{\it Dark monopoles} may constitute a significant fraction of {\it dark sector}. Namely magnetic dipoles can be coupled to a massive
{\it dark photon} and the kinetic mixing term between the two photons can gives the aforementioned dipoles a small coupling to the {\it visible photon} \cite{ter19}.
In the scenario  where the {\it kinetic mixing term} between {\it visible} and {\it dark photon} sectors appears,
magnetic monopoles of broken {\it dark symmetry} group
may emerge in the {\it visible sector} as confined milli-magnetically charged objects. Studies of the interaction between magnetic galactic field and milli-charged particles 
have been given in \cite{gra22}. Among all, it was revealed that for a wide range of magnetic monopole and {it dark photon} masses, the energy extraction from galactic magnetic field could
put the leading constraints on the model in question, providing that the milli-charged magnetic monopoles comprised all of the {\it dark matter} hallo.

The other investigations in the
aforementioned 
direction were given in \cite{ven21}, where the Klein-Kaluza monopolium (if existent) might be a candidate for a primordial {\it dark sector} component.
In \cite{yan23} the possibilities of {\it dark monopoles}
serving as {\it dark matter} and {\it dark sector} first-order phase transition remnants were elaborated. It turned out that heavy monopoles produced from the first-order
high-scale phase transitions might be responsible for producing weak gravitational waves. They can contribute as the considerable component of {\it dark matter} relic densities.

The further studies, conducted 
in Ref. \cite{hir21} have revealed how {\it dark topological defects} affect the Standard Model (SM), paying attention to the influence of
$\alpha$-coupling constant between {\it visible } and {\it dark sectors}. Among all the {\it dark monopole}, {\it dark strings} and the so-called {\it dark beads} solutions when the 
symmetry $SU(2) \to U(1) \to Z_2$, were elaborated. Moreover, interactions of electrical and magnetic charges and topological defects were treated in \cite{chi23}, where in 
comparison to \cite{hir21},
the magnetic aspects of the problem in questions like dyons and dyonic beads were considered.

Having in mind the possible importance of the {\it dark matter monopole} for the physics of our Universe,
in this section we shall consider magnetic monopoles attributed to both {\it visible} and {\it dark sectors} and the adequate electric charges of the aforementioned sectors.
The charges will be situated at the distance, say, $c$ from magnetic monopoles.

The magnetic monopoles connected with the {\it visible} and {\it dark sectors} are described by the relations
\ben
 \tB_{(\tE)} ^m= \frac{g_{\tE}~ x^m}{{ r}^3} = \frac{\sqrt{2 - \alpha}}{2} \frac{(g_E - g_B)~x^m}{r^3},\\
 \tB_{(\tB)} ^m= \frac{g_{\tB}~ x^m}{{ r}^3} = \frac{\sqrt{2 + \alpha}}{2} \frac{(g_E + g_B)~x^m}{r^3}.
\een
We
 assume the form of $g_\tE$ and $g_\tB$, to be consistent with the definitions of the charges given by (\ref{cA})-(\ref{cB})
  in the {\it dark photon} theory.

On the other hand, the electric fields have the following forms:
\ben
\tE_{(\tE)}^k = \frac{\te~ (x^k- x^{'k})}{\mid {\vec r} -{\vec r'}\mid^3} = \frac{\sqrt{2 - \alpha}}{2} \frac{(e - e_d)~(x^k- x^{'k})}{\mid {\vec r} -{\vec r'}\mid^3},\\
\tE_{(\tB)}^k = \frac{\te_d~ (x^k- x^{'k})}{\mid {\vec r} -{\vec r'}\mid^3} = \frac{\sqrt{2 + \alpha}}{2} \frac{(e + e_d)~(x^k- x^{'k})}{\mid {\vec r} -{\vec r'}\mid^3}.
\een
All the above lead to the form of
 the azimuthal components of both sectors gauge fields, provided by the following:
\ben \label{mon}
A_{\phi (\tE)} =  \frac{\sqrt{2 - \alpha}}{2} (g_E - g_B) (1 - \cos \theta ),\\
A_{\phi (\tB)} =  \frac{\sqrt{2 + \alpha}}{2} (g_E + g_B) (1 - \cos \theta ).
\een
The general form angular momentum stored in the {\it visible} and {\it hidden sectors} fields is given by the relation
\be \label{jz}
J_a = - \frac{1}{4 \pi} \int d^3x~\Big[ \ep_{abc} x^b \Big( \ep^{cij} {\tE_{(\tE) i}}{\tB_{(\tE) j }} \Big)+
\ep_{abc} x^b \Big( \ep^{cij} {\tE_{(\tB) i}}{\tB_{(\tB) j }} \Big)
\Big].
 \ee
 Using the relations
 for magnetic fields for both sectors given by the following equations:
 \be
 \tB_{(\tF)_j}  = \frac{1}{2} \ep_{jkl} \sqrt{g}~\tF^{kl}, \qquad    \tB_{(\tB)_j}  = \frac{1}{2} \ep_{jkl} \sqrt{g}~ \tB^{kl},
 \ee
the relation (\ref{jz}), for the its $z$-component, may be rewritten in the form as
\be \label{jzz}
J_z =  - \frac{1}{4 \pi} \int dr d\theta d\phi ~\tF_{\phi m}  \cE_{(\tE)}^m 
- \frac{1}{4 \pi} \int dr d\theta d \phi ~\tB_{\phi m} \cE_{(\tB)}^m,
\ee
where we set $\cE_{(\tE)}^a = \sqrt{g} \tE_{(\tE)}^a$, and $\cE_{(\tB)}^a = \sqrt{g} \tE_{(\tE)}^a$.

In what follows one can point out that $A_{\phi (\tE)}$ and $A_{\phi (\tB)}$ depend on the $\theta$-angle, while the electric field densities  on
$r$ and $\theta$-coordinates. These facts enable us to rewrite (\ref{jzz}) in the following form:
\be
J_z =  - \frac{1}{4 \pi} \int dr d\theta d\phi ~\Bigg[ \p_\theta A_{\phi (\tE)} \cE^\theta_{(\tE)} +  \p_\theta A_{\phi (\tB)} \cE^\theta_{(\tB)} \Big].
\ee
Further integration by parts and the remark that $\cE^\theta_{(\tE)}$ and $\cE^\theta_{(\tB)}$ vanish at $\theta=0$ and $\theta = \pi$, authorize that
\be
J_z =  - \frac{1}{4 \pi} \int dr d\theta d\phi ~\Bigg[ A_{\phi (\tE)}~ \p_\theta \cE^\theta_{(\tE)} +  A_{\phi (\tB)}~ \p_\theta \cE^\theta_{(\tB)} \Big].
\ee
In the next step we take into account the explicit forms of $\phi$-components of the potentials, as well as, 
\ben
\p_r \cE_{(\tE)} + \p_\theta \cE^\theta_{(\tE)} &=& 4 \pi ~\te~\delta(r-c) \delta^{(2)}(\theta, \phi),\\
\p_r \cE_{(\tB)} + \p_\theta \cE^\theta_{(\tB)} &=& 4 \pi ~\te_d~\delta(r-c) \delta^{(2)}(\theta, \phi),
\een
where $\delta^{(2)}(\theta, \phi)$ \cite{bun07} depicts the $\delta$-function density on the sphere with a support at the northern pole,
namely $\int f(\theta,~\phi) \delta^{(2)}(\theta,~ \phi) = f(\theta = 0).$
All the above,
lead to the expression for $J_z$ in a region of spacetime bounded by the two spheres of adequate radii, $r_1$ and $r_2$, i.e.,
\ben \label{str}
J_z &=& \int_{r_1}^{r_2} dr~\frac{d}{dr} \Big[ g_{\tE} ~\te~\Theta(r-c) - \frac{g_{\tE}}{2} \int_0^{\pi} d\theta~\cos \theta ~\cE^r_{(\tE)}(r,~\theta) \Big] \\ \nonumber
&+& \int_{r_1}^{r_2} dr~\frac{d}{dr} \Big[ g_{\tB}~ \te_d~\Theta(r-c) - \frac{g_{\tB}}{2} \int_0^{\pi} d\theta~\cos \theta ~\cE^r_{(\tB)}(r,~\theta) \Big], 
\een
where $\Theta(r)$ stands for the Heaviside step function.

As was pointed out in Ref. \cite{bun07} the form of the above equation allows us to present the angular momentum contained between two spheres $r_1,~r_2$ as the difference
$J_z = \Phi (r_2) - \Phi(r_1)$, where is give by the expression in the brackets of equation (\ref{str}) and note that in particular we have that $\Phi(0)=0$. Just we can conclude that
$J_z = \Phi(r)$. From the form of $\cE_{(\tE)(\tB)}^r = \te (\te)_d ~\sin \theta$ we have that
\be
J_z = \Phi(\infty) =  \te~g_{\tE} + \te_d~g_{\tB}.
\label{jinfty}
\ee
If we demand that $J_z$ should be quantized to half integer one arrives at
\be
\frac{2}{\hbar} \Big( \te ~g_{\tE} + \te_d~g_{\tB} \Big) \in Z.
\label{qcon}
\ee
Let us consider the special cases of the above relation:\\
1. we have no {\it dark matter}, i.e., $\alpha =0,~e_d =0, ~g_B =0$, then the equation (\ref{qcon}) is given by
\be
\frac{2}{\hbar} e~g_E  \in Z,
\ee
one obtains the Dirac-like quantization conditions, known in Maxwell theory.\\
2. we set $e_d =0$, one arrives at the following generalization of Dirac quantization condition, taking into account {\it dark matter} sector
\be
\frac{2}{\hbar} e~\Big(g_E + \frac{\alpha}{2} g_B \Big)  \in Z.
\ee

\section{Possible observational consequences of dark monopole existence}
Up to now the detection of monopoles constitutes a big challenge for the experimental physics. A variety of sophisticated techniques are implemented in the search
for this elusive object which existence has been foreseen by various unified gauge theories, having very general consequences of the unification of the fundamental interactions and history 
and the very Early Universe physics (see, e.g. \cite{pre84}).

In one of the possible experimental verification of the derived quantization condition, we shall use Aharonov-Bohm-like (AB) attitude \cite{aha59}.
Namely, one shall look for the phase difference in the closed paths, of the beam of $\te$ and $\te_d$ charges circling magnetic monopoles, given by relations (\ref{mon}).
In what follows, for $\theta$ we set $\pi/2$.  

In order to find the phase in AB-like experiment, for magnetic gauge monopoles $A_{\phi (\tE)},~A_{\phi (\tB)} $, one implements the path integral formulation \cite{fel}, and firstly
write the action integral of the charge $\te,~\te_d$ interactions with the adequate gauge fields.  
\be
S = S_{\tA} + S_{\tB} = \te \int_{\Gamma} \tA_\mu dx^\mu + \te_d \int_{\Gamma} \tB_\mu dx^\mu,
\ee
where $\Gamma$ is the integral path.
On the other hand, phases are given by
\be
f = f_{\tA} + f_{\tB} = \frac{\te}{\hbar}  \int_{\Gamma} \tA_\mu dx^\mu + \frac{\te_d}{\hbar} \int_{\Gamma} \tB_\mu dx^\mu.
\ee
 In the experiment under inspection we measure the phase difference
$\Delta f = \Delta f_{\tA} + \Delta f_{\tB} $ in closed paths.  Consequently, the phase difference is provided by
\ben
\Delta f &=& \Delta f_{\tA} + \Delta f_{\tB} \\ \nonumber
&=& \frac{\te}{\hbar} \Big( \int_{\Gamma_1} \tA_\mu dx^\mu - \int_{\Gamma_2} \tA_\mu dx^\mu \Big)\\ \nonumber
&+& \frac{\te_d}{\hbar} \Big( \int_{\Gamma_1} \tB_\mu dx^\mu - \int_{\Gamma_2} \tB_\mu dx^\mu \Big)\\ \nonumber
&=&
\frac{\te}{\hbar} \oint A_{\phi (\tE)} d\phi + \frac{\te_d}{\hbar} \oint A_{\phi (\tB)} d\phi.
\een
Taking into account that $\phi$ goes from $-\pi/2$ to $\pi/2$, and from $3 \pi/2$ to $\pi/2$ in the clockwise direction,
we have that $\oint d \phi = 2 \pi$. It enables to find that
\be
\Delta f = \frac{2 \pi}{\hbar} \Big( \te ~\tg_E + \te_d ~\tg_B \Big).
\label{ff}
\ee
The similar discussion of the equation (\ref{ff}), like the one after the relation (\ref{qcon}), reveals that we obtain the modification of the phase shift comparing to the Dirac monopole case.

Because of the fact that the shifts in AB-like experiments are small, one should wait for the improvement in technology. The last generation of AB experiments, using modified 
electron microscopes with $\mu m$ resolution and long exposure time, have sensitivities in the range of $10^{-2}-10^{-3}$ \cite{abexp}.

Of course, the other observational techniques, which are implemented in the searches of monopole in the visible sector, can be applied \cite{raj16}, i.e.,
SQUID-technology, accelerator experiments, solid state experiments.

Now, let us have a glimpse at the possible effect of {\it dark matter monopole} on the current through the condensed matter SQUID, i.e., 
we imagine the {\it dark matter monopole} travel across one of its junctions \cite{ann09}.
To find the changes of the flux $\Phi$ induced by the passage of the {\it dark monopole}, let us recall the duality relation between electric and magnetic parts of
respectively $\tF_{\alpha \beta}$ and
$\tB_{\alpha \beta}$, as well as their equations of motion, i.e., 
\ben
\na^{\alpha} \tF_{\alpha \beta} &=& {\tilde j}(\te), \qquad \na^{\alpha}  \ast \tF_{\alpha \beta} = - {\tilde j}(\tg_E), \\
\na^{\alpha} \tB_{\alpha \beta} &=& {\tilde j}(\te_d), \qquad \na^{\alpha}  \ast \tB_{\alpha \beta} = - {\tilde j}(\tg_B), 
\een
where the dual tensors are given by $ \ast \tF_{\alpha \beta} = 1/2 \ep_{\alpha \beta \ga \delta} \tF^{\ga \delta}$ and $ \ast \tB_{\alpha \beta} = 1/2 \ep_{\alpha \beta \ga \delta} \tB^{\ga \delta}$.

Thus the corresponding magnetic fluxes 
are provided by 
\ben
\Phi_{\tA} &=& \frac{\sqrt{2 - \alpha}}{2} \Bigg[ \frac{2 \pi ~a}{g_E} - \frac{2 \pi~b}{g_B} \Bigg],\\
\Phi_{\tB} &=& \frac{\sqrt{2 + \alpha}}{2} \Bigg[ \frac{2 \pi ~a}{g_E} + \frac{2 \pi~b}{g_B} \Bigg],
\een
while $a,~b \in {\bf Z}$.
Thus the total magnetic flux encompassed  by the SQUID device can be written as $\Phi_{\tA} + \Phi_{\tB} $. 
 Taking into account the fact hat the coupling constant $\alpha \ll 1$, one can rewrite the sum in the forms as
\be
\Phi_{\tA} + \Phi_{\tB}  \simeq 
\frac{2 \sqrt{2} \pi}{e} \Bigg[ a + \frac{\alpha}{4} \Big( \frac{g_E}{g_B} \Big) ~b\Bigg].
\ee
If one assumes that $a=b$, we obtain the following:
\be 
\Phi_{\tA} + \Phi_{\tB}  \simeq  \Phi_A \Bigg[ 1 + \frac{\alpha}{4} \Big( \frac{g_E}{g_B} \Big) \Bigg],
\label{eff-fux}
\ee
where $\Phi_A$ is the magnetic flux bounded with $U(1)$-visible sector monopole. 

On the basis of the equation (\ref{eff-fux}) and assuming that $\frac{g_E}{g_B} = 1$, we argue that due to the {\it kinetic mixing} (with the coupling $\alpha$) the effective 
magnetic field and thus the flux through the device would  change from the value $\Phi$ to the new value $(1+\alpha/4)\Phi$ when the {\it dark monopole} is crossing the SQUID, so obviously the current in the device would also change.


\section{Dark photon electric charge in the vicinity of magnetic black hole with dark sector fields}
In this section we shall pay attention to the problem of the existence of generalized electrical charges $\te$ and $\te_d$, connected with both studied sectors,
in the vicinity of magnetically charged black hole. The magnetic charges will also be bounded with {\it visible}, $g_E$, and {\it dark sectors}, $g_B$.

In order to investigate the spin-up process of black hole caused by {\it dark monopole}, we shall consider the line elements of static and stationary axisymmetric (rotating)
black holes in Einstein- {\it dark photon} gravity derived by the present author in Refs. \cite{kic19} and \cite{rog24}. The crucial point in the final conclusion, will constitute the uniqueness theorem for stationary axisymmetric {\it dark photon} black hole also found by the author in \cite{rog24}.

\subsection{Black holes in dark photon theory}
\subsubsection{Static black hole solution with dark matter sector}
The line element of static spherically symmetric black hole in the theory in question, with
 timelike Killing vector $(\frac{\p}{ \p t})_a$
orthogonal to the spacelike hypersurface in this spacetime,
 is given by \cite{kic19}
\be
ds^2 = -\Big( 1 - \frac{2m}{r} + \frac{g_E^2 + \alpha g_E g_B + g_B^2}{r^2} \Big) dt^2 +\frac{dr^2}{\Big( 1 - \frac{2m}{r} + \frac{g_E^2 + \alpha g_E g_B + g_B^2}{r^2} \Big)}
+ r^2 \Big(d\theta^2 + \sin^2 \theta d\phi^2 \Big),
\label{static}
\ee
where we have assumed that one has only magnetically charged static black hole.

\subsubsection{Rotating black hole with dark matter sector }
The solution of Einstein-Maxwell {\it dark photon} theory o gravity describing the stationary axisymmetric black hole was obtained in Ref. \cite{rog24}.
It has been also proved that stationary axisymmetric solutions of the aforementioned theory of gravity subject to the same boundary and regularity conditions
(Kerr-Newman-like {\it dark photon} black holes) possessing regular event horizon, are the only stationary axisymmetric black hole solutions in Einstein-Maxwell {\it dark photon } gravity
(the uniqueness theorem for Kerr-Newman-like black holes in Einstein-Maxwell {\it dark photon} theory).

It happens that if we define potentials, in Ernst-like construction leading to this solution,
 in the following forms:
\be
\Phi_{(\tF)} = \frac{Q_{(\tF)}}{\xi+1}, \qquad \Phi_{(\tB)} = \frac{Q_{(\tB)}}{\xi+1},
\ee
as well as their components
\ben
Q_{(\tF)} &=& \frac{\sqrt{2-\alpha}}{2} \Big[ \Big( e - e_d \Big) + i \Big(g_E - g_B\Big)\Big],\\
Q_{(\tB)} &=& \frac{\sqrt{2+\alpha}}{2} \Big[ \Big( e + e_d \Big) + i \Big(g_E + g_B\Big)\Big],
\een
then the Kerr-Newman like black hole in {\it dark photon} theory, with electric and magnetic charges under both $U(1)$-gauge groups is written in
Boyer-Lindquist coordinates, as follows:
\ben \label{rot bh}
ds^2 = &-& \frac{\Delta}{r^2} \Big[ dt - a~\sin^2 \theta d \phi \Big]^2 + \frac{\sin^2 \theta}{\rho^2} \Big[ \Big(r^2 + a^2 \Big) d\phi - a~dt \Big]^2 \\ \nonumber
&+& \frac{\rho^2}{\Delta} dr^2 + \rho^2~d \theta^2,
\een
where we denote by $\Delta$ and $\rho$, respectively
\be
\Delta = r^2 - 2 mr + e^2 + e_d^2 + \alpha~ ee_d +g_E^2 + g_B^2 + \alpha ~g_E g_B, \qquad \rho^2 = r^2 + a^2~\cos^2 \theta.
\ee
On the other hand, 
the asymptotic form of stationary axisymmetric black hole with both sectors magnetic charges (linearized in electrical charges) is of the form given by
\ben \label{rot}
ds^2 = - \frac{\tilde \Delta}{r^2} dt^2 &+& r^2 \sin^2 \theta d\phi^2 + \frac{2a \sin^2 \theta}{r^2} \Big[ -2 m r + g_E^2 + g_B^2 + \alpha~ g_E g_B \Big] dt d\phi \\ \nonumber
&+ & \frac{r^2}{\tilde \Delta} dr^2 + r^2~d\theta^2,
\een
where $\tilde \Delta = r^2 - 2 mr + g_E^2 + g_B^2 + \alpha ~g_E g_B$.
It can be also noticed that the line element reduces to the static spherically symmetric one, described by the relation (\ref{static}), when angular momentum per unit mass $a$ is equal to zero.

\subsection{Rotating black hole and monopoles}
First we consider the situation when electrically charged particles will be placed at rest, at the specified distance, in the spacetime of black hole magnetically charged under both 
$U(1)$ symmetry groups.

Because of the fact that a global conserved charge is connected with a gauge symmetry, the total angular momentum can be described as a surface 
integral at infinity. The surface integral in question is specified by the demand that the relative generator ought to possess a well defined functional 
derivative  describing $J_z$ \cite{bun07}. All the above implicate that the aforementioned generator can be provided by
\be
G = \int d^3x~\chi^m ~\cH_m + J_z,
\ee
where $\chi^m$ is an arbitrary function of spatial coordinates, having the property that $\chi^m \rightarrow 1$ as $r \rightarrow \infty$ \cite{bun07}.


On the other hand, the Hamiltonian generator in the case under consideration, is given by
\be
\cH_\phi = -2 \na_k \pi^k_\phi - \frac{1}{4 \pi}\tF_{\phi m}  \cE_{(\tE)}^m - \frac{1}{4 \pi}~\tB_{\phi m} \cE_{(\tB)}^m,
\ee

while the variation of $G$ is provided by
\be
\delta G = \int_{vol} A - 2 \int_{S^2_\infty} d \theta d\phi~\delta \pi^r{}{}_\phi,
\ee
where $\int_{vol} A$ is a volume integral.
Thus as in Ref. \cite{bun07}, we can equate the last term as the $z$-component of angular momentum, i.e., 
\be
J_z = 2 \int_{S^2_\infty} d \theta d\phi~\pi^r{}{}_\phi.
\ee
To find the value of $J_z$, we integrate the constraint equation $\cH_\phi =0$, over a two-sphere (from $r_+$ to infinity). It leads to the relation
\be
J_z = \Phi(\infty) - \Phi(r_+) + 2 \int_{S^2_{r_+}} d \theta d\phi~ \pi^r{}{}_\phi.
\label{spin}
\ee
As was mentioned in Ref. \cite{bun07}, one can calculate the above integral for $c \gg r_+$ (is the same as letting $r_+ \rightarrow \infty$, and keeping $c$ finite). The spacetime
becomes flat and the domain of integration for each of the integrals at $r_+$ is a two-sphere of vanishing radius. It makes the integral
vanish, because the integrand is regular, in the sense that density of positive weight disappears at the origin of coordinate system. So for any radial coordinate value
$c$, we reach to the conclusion that 
\be 
J_z = \Phi(\infty) = \te~g_{\tE} + \te_d~g_{\tB}.
\ee
This fact authorizes that the total angular momentum comprises the two components. One is bounded with angular momentum stored in 
two $U(1)$-gauge fields, while the spin of black hole is govern by the second term of the equation (\ref{spin}).\\

In the next step we shall find the exact form of the spin of black hole given by the second term in relation (\ref{spin}).
Namely, having in mind the relation for $ \pi^r{}{}_\phi$ given by
\be
 \pi^r{}{}_\phi = - \frac{1}{16 \pi} K_{\phi r}~ g^{rr} \sqrt{g},
 \ee
 and finding the exact form of the exterior curvature 
 \be
 K_{\phi r} = \frac{1}{2 N} \Big( \na_r N_\phi + \na_\phi N_r \Big),
 \ee
 where $N$ is the lapse function, and $N_{\phi,r}$ are the adequate components of shift functions for the line element (\ref{rot}) on the slices $t = constant$,
 one arrives at $r_+$ the following expression:
\be
2 \int_{S^2_{r_+}} d \theta d\phi~ \pi^r{}{}_\phi = - ma + \frac{2a}{3 r_+} \Big(g_E^2 + g_B^2 + \alpha ~g_E g_B \Big).
\ee
We take into account the situation that point charged particles are moving toward the black hole event horizon. One can imagine that it crosses the event horizon and finds itself inside.
When the charges are far away outside the event horizon of black hole, the angular momentum is stored in both two gauge fields and we have
a non-rotating black hole static, spherically symmetric, given by the line element (\ref{static}). As the charges are taken closer to black hole, the angular momentum
in the fields in question begins to continuously transfer towards black hole. The closer are the charges $\te$ and $\te_d$, the faster black hole spins.
When the charges have reached the event horizon, the transfer ends and black hole is rotating at the required rate (it happens without black hole jolt).

Thus we encounter the situation when the field equations with magnetic and electric charges are in the spacetime with two Killing vectors, i.e.,
$(\frac{\p}{ \p t})_a$ and $(\frac{\p}{ \p \phi})_m$. 

Due to the uniqueness theorem for stationary axisymmetric black hole with {\it dark matter} proved in Ref. \cite{rog24}, the exterior solution is given by the metric (\ref{rot})
(when linearized in electrical charges), with an angular momentum/per unit mass $a$ is provided by the relation
\be
- ma =  \te~g_{\tE} + \te_d~g_{\tB}.
\label{aaa}
\ee
In obedience to the uniqueness theorem for stationary axisymmetric black holes in Einstein-Maxwell {\it dark photon} gravity, the angular
momentum bounded with the charges $\te,~\te_d$ - monopoles $g_{\tE},~g_{\tB}$ system, mislays all its exotic features and the observer from the outside
region becomes cognisant of a common rotation of the black hole in question.

\section{Possible observational consequences}
\subsection{Rotation parameter of black hole}
As was seen from the above discussion, the crucial point of our investigation is the conclusion that the mixture of Maxwell and {\it dark } charges, as well as,
magnetic monopoles connected with both sectors, loose their exotic properties giving rise to the rotation of black hole.

Let us have a closer look at the relation (\ref{aaa}). Having in mind that $\alpha$-coupling has small vale (see discussion after equation (\ref{ac dm}))), 
and putting $e_d =0$,
one can find that
\be
- ma = e \Big( g_E + \frac{\alpha}{2} g_B \Big).
\ee
Thus, we get the difference in the measurements of $-ma$,
comparing to the Maxwell case, which states that
\be
- ma =  e~g_E,
\label{max p}
\ee
which can be possibly detected by the future sophisticated experiments, which will look for black hole rotation parameter which will be greater than (\ref{max p}).\\
Moreover, if we assume that $g_E = g_B$, which is the plausible assumption in the light of the considerations given in Ref. \cite{ter19}, then we arrive at
the following:
\be
-ma = e~g_E \Big( 1 + \frac{\alpha}{2} \Big).
\ee
Thus the rotation parameter takes the value greater than in Maxwell case.
Concluding, looking for the rotating black holes, with the anomalous rotation parameter, can constitute the possible imprint of {\it dark monopole} trace.
However because of the small value of the $\alpha$-parameter the future very sophisticated experimental method ought to be applied. The difference between these 
quantities will be proportional to $\alpha~g_E$,

However, the lack of experimental detections of {\it dark monopole}/monopole may be explained by such black hole spin-up process, where these exotic
objects end their lives in the black hole interiors.

As far as the black holes with the magnetic charges, like give by the line elements (\ref{static}) and (\ref{rot}), one can look for the modifications of charge particle trajectories 
around them (like e.g., \cite{rog25a}), synchrotron radiation \cite{rog25b}, or the possible modification of the black hole shadow.

On the other had, one can obtain the upper limit on {\it dark monopole} flux from various astrophysical observations \cite{gia03}, like
upper limit from mass density of the Universe, limit from galactic and intergalactic magnetic fields, limit from pulsars physics.


\section{Conclusions}
In our paper we have considered the problem of electrical charges, connected with {\it visible} and {\it hidden sectors} in the model of {\it dark photon},
in the vicinity of magnetically charged static spherically symmetric black hole. The presence of charges will constitute
the exterior perturbations of magnetically charged black hole.

As the charges move in, the {\it hair} on the black hole spacetime will fade away. It results in progressively better approximation of the actual metric by the line element (\ref{rot}).
It will be exact when charges in question reach the black hole event horizon.

This situation can be explained
due to the uniqueness theorem for these kind of black holes proved in \cite{rog24}, when the charges will be devoured by the black hole, the angular momentum connected with
charges-monopoles system will be spotted from the outside as a rotation of the black hole in question (black hole will has no hair).

From the point of view of the duality of both $U(1)$-gauge theories (see the equations of motion for the underlying theory (\ref{fb})),
one can make a static black hole a rotating one, by throwing radially into its event horizon a magnetic monopole (the situation is equivalent to placing
magnetic monopole outside an electrically charged black hole).

Additionally one should mention that, the same process as described in \cite{bun07} can be conceivable, i.e., we can make black hole rotate by delivering radially into it magnetic poles.
Namely, the sequence of radial throwing of $\te,~\te_d$ and $g_{\tE},~g_{\tB}$ to the northern pole of static black hole, and 
$-\te,~-\te_d$ and $-g_{\tE},~-g_{\tB}$, to the southern pole, will result in obtaining rotating black hole.

Of course, one can speculate that it is conceivable that during the history of our Universe some monopoles have ended their lives in the black holes, giving rise to
black hole rotation.

\acknowledgments 
MR was partially supported by Grant No. 2022/45/B/ST2/00013 of the National Science Center, Poland.






\begin{thebibliography}{99}

%
\def\cmp#1#2#3#4{\emph{#4}, \emph{ Commun. Math. Phys.} {\bf #1} (#3) #2}
\def\lmp#1#2#3#4{\emph{#4}, \emph{ Lett. Math. Phys.} {\bf #1} (#3) #2}
\def\hpa#1#2#3#4{\emph{#4}, \emph{ Hell. Phys. Acta} {\bf #1} (#3) #2}
\def\grg#1#2#3#4{\emph{#4}, \emph{ Gen. Rel. Grav.} {\bf #1} (#3) #2}
\def\pr#1#2#3#4{\emph{#4}, \emph{ Phys. Rev.} {\bf #1} (#3) #2}
\def\prl#1#2#3#4{\emph{#4}, \emph{ Phys. Rev. Lett.} {\bf #1}, #2 (#3)}
\def\prd#1#2#3#4{\emph{#4}, \emph{ Phys. Rev. D} {\bf #1}, #2 (#3)}

\def\prb#1#2#3#4{\emph{#4}, \emph{ Phys. Rev. B} {\bf #1}, #2 (#3) }
\def\prx#1#2#3#4{\emph{#4}, \emph{ Phys. Rev. X} {\bf #1} (#3) #2}
\def\pl#1#2#3#4{\emph{#4}, \emph{ Phys. Lett.} {\bf #1} (#3) #2}
\def\pla#1#2#3#4{\emph{#4}, \emph{ Phys. Lett. A} {\bf #1} (#3) #2 }
\def\plb#1#2#3#4{\emph{#4}, \emph{ Phys. Lett. B} {\bf #1}, #2 (#3)}
\def\prep#1#2#3#4{\emph{#4}, \emph{ Phys. Reports} {\bf #1}, #2 (#3)}
\def\phys#1#2#3#4{\emph{#4}, \emph{ Physica} {\bf #1} (#3) #2}
\def\jcp#1#2#3#4{\emph{#4}, \emph{ J. Comput. Phys.} {\bf #1} (#3) #2}
\def\jmp#1#2#3#4{\emph{#4}, \emph{ J. Math. Phys.} {\bf #1} (#3) #2}
\def\jpm#1#2#3#4{\emph{#4}, \emph{ J. Phys. A: Math. Gen.} {\bf #1} (#3) #2}
\def\cpr#1#2#3#4{\emph{#4}, \emph{ Computer Phys. Rept.} {\bf #1} (#3) #2}
\def\cqg#1#2#3#4{\emph{#4}, \emph{ Class. Quant. Grav.} {\bf #1} (#3) #2}
\def\cma#1#2#3#4{\emph{#4}, \emph{ Computers Math. Applic.} {\bf #1} (#3) #2}
\def\mc#1#2#3#4{\emph{#4}, \emph{ Math. Compt.} {\bf #1} (#3) #2}
\def\apj#1#2#3#4{\emph{#4}, \emph{ Astrophys. J.} {\bf #1} (#3) #2}
\def\apjs#1#2#3#4{\emph{#4}, \emph{ Astrophys. J. Suppl.} {\bf #1} (#3) #2}
\def\apjl#1#2#3#4{\emph{#4}, \emph{ Astrophys. J. Lett.} {\bf #1} (#3) #2}
\def\acta#1#2#3#4{\emph{#4}, \emph{ Acta Astronomica} {\bf #1} (#3) #2}
\def\apl#1#2#3#4{\emph{#4}, \emph{ Ann. Physik. (Leipzig)} {\bf #1} (#3) #2}
\def\amjp#1#2#3#4{\emph{#4}, \emph{Am. J. Phys.} {\bf #1} (#3) #2}
\def\anp#1#2#3#4{\emph{#4}, \emph{ Ann. Phys.} {\bf #1} (#3) #2}
\def\sa#1#2#3#4{\emph{#4}, \emph{ Sov. Astro.} {\bf #1} (#3) #2}
\def\sia#1#2#3#4{\emph{#4}, \emph{ SIAM J. Sci. Statist. Comput.} {\bf #1} (#3) #2}
\def\aa#1#2#3#4{\emph{#4}, \emph{ Astron. Astrophys.} {\bf #1} (#3) #2}
\def\mnras#1#2#3#4{\emph{#4}, \emph{ Mon. Not. R. Astr. Soc.} {\bf #1} (#3) #2}
\def\npb#1#2#3#4{\emph{#4}, \emph{ Nucl. Phys. B} {\bf #1}, #2 (#3)}
\def\npa#1#2#3#4{\emph{#4}, \emph{ Nucl. Phys. A} {\bf #1} (#3) #2}

\def\prsla#1#2#3#4{\emph{#4}, \emph{ Proc. R. Soc. London, Ser. A} {\bf #1} (#3) #2}
\def\jhep#1#2#3#4{\emph{#4}, \emph{ JHEP} {\bf #1} (#2) #3}
\def\jcap#1#2#3#4{\emph{#4}, \emph{ JCAP} {\bf #1} (#2) #3}

\def\nuca#1#2#3#4{\emph{#4}, \emph{ Nuovo Cimento A } {\bf #1} (#3) #2}
\def\nucb#1#2#3#4{\emph{#4}, \emph{ Nuovo Cimento B } {\bf #1} (#3) #2}
\def\ijmp#1#2#3#4{\emph{#4}, \emph{ Int. J. Mod. Phys. D} {\bf #1} (#3) #2}
\def\atmp#1#2#3#4{\emph{#4}, \emph{ Adv. Theor. Math. Phys.} {\bf #1} (#3) #2}
\def\ptps#1#2#3#4{\emph{#4}, \emph{ Prog. Theor. Phys. Suppl.} {\bf #1} (#3) #2}
\def\ptp#1#2#3#4{\emph{#4}, \emph{ Prog. Theor. Phys.} {\bf #1} (#3) #2}
\def\lmp#1#2#3#4{\emph{#4}, \emph{ Lett. Math. Phys.} {\bf #1} (#3) #2}
\def\cpam#1#2#3#4{\emph{#4}, \emph{ Comm. Pure Appl. Math.}  {\bf #1} (#3) #2}
\def\adv#1#2#3#4{\emph{#4}, \emph{ Adv. Phys.}  {\bf #1} (#3) #2}
\def\zh#1#2#3#4{\emph{#4}, \emph{ Zh. Eksp. Teor. Fiz.}  {\bf #1} (#3) #2}
\def\mplb#1#2#3#4{\emph{#4}, \emph{ Mod. Phys. Lett. B} {\bf #1}, #2 (#3)}
\def\mpla#1#2#3#4{\emph{#4}, \emph{ Mod. Phys. Lett. A} {\bf #1}, #2 (#3)}


\def\jams#1#2#3#4{\emph{#4}, \emph{ J. Austral. Math. Soc. B} {\bf #1} (#3) #2}
\def\appa#1#2#3#4{\emph{#4}, \emph{ Acta Phys. Polonica A} {\bf #1} (#3) #2}
\def\appb#1#2#3#4{\emph{#4}, \emph{ Acta Phys. Polonica B} {\bf #1} (#3) #2}

\def\nat#1#2#3#4{\emph{#4}, \emph{Nature} {\bf #1} #2 (#3)}
\def\natcom#1#2#3#4{\emph{#4}, \emph{Nature Commun.} {\bf #1} (#3) #2}
\def\natphys#1#2#3#4{\emph{#4}, \emph{Nature Physics} {\bf #1} (#3) #2}
\def\natmat#1#2#3#4{\emph{#4}, \emph{Nature Mat.} {\bf #1} (#3) #2}


\def\science#1#2#3#4{\emph{#4}, \emph{Science} {\bf #1} (#3) #2}
\def\sciadv#1#2#3#4{\emph{#4}, \emph{Sci. Adv.} {\bf #1} (#3) #2}

\def\arcmp#1#2#3#4{\emph{#4}, \emph{Annual Rev. of Cond. Matter Physics} {\bf #1} (#3) #2}
\def\zphys#1#2#3#4{\emph{#4}, \emph{Z. Phys.} {\bf #1}, (#3) #2}
\def\ncs#1#2#3#4{\emph{#4}, \emph{Nuovo Cimento Suppl.} {\bf #1} (#3) #2}
\def\physb#1#2#3#4{\emph{#4}, \emph{Physica B} {\bf #1}, (#3) #2}
\def\jpcm#1#2#3#4{\emph{#4}, \emph{J. Phys.: Condens. Matter } {\bf #1} (#3) #2}
\def\pnas#1#2#3#4{\emph{#4}, \emph{Proc. Nat. Academy Sciences} {\bf #1} (#3) #2}
\def\sssr#1#2#3#4{\emph{#4}, \emph{Izv. Akad Nauk SSSR, ser. fiz.} {\bf #1} (#3) #2}
\def\jpg#1#2#3#4{\emph{#4}, \emph{ J. Phys. G} {\bf #1} (#3) #2}
\def\chinpb#1#2#3#4{\emph{#4}, \emph{Chin. Phys. B} {\bf #1} (#3) #2}
\def\njp#1#2#3#4{\emph{#4}, \emph{ New J. Phys.} {\bf #1} (#3) #2}
\def\frontphys#1#2#3#4{\emph{#4}, \emph{ Front. Phys.} {\bf #1} (#3) #2}
\def\epl#1#2#3#4{\emph{#4}, \emph{ EPL} {\bf #1} (#3) #2}
\def\rmp#1#2#3#4{\emph{#4}, \emph{ Rev. Mod. Phys.} {\bf #1}, #2 (#3)}
\def\rpp#1#2#3#4{\emph{#4}, \emph{ Rep. Prog. Phys.} {\bf #1}, #2 (#3)}

\def\hepph#1#2{{ hep-ph }{#1} (#2)}
\def\arxiv#1#2#3{\emph{#3},{ arXiv }{#1} (#2)}
\def\hepth#1#2{{ hep-th }{#1} (#2)}
\def\grqc#1#2{{ gr-qc }{#1} (#2)}
\def\ibid#1#2#3#4{\emph{#4}, {\it ibid.} {\bf #1} (#3) #2}
\def\conphy#1#2#3#4{\emph{#4}, \emph{Contemporary Physics} {\bf #1}, (#3) #2}
\def\ppnp#1#2#3#4{\emph{#4}, \emph{ Prog. Part. Nucl. Phys} {\bf #1} (#3) #2}
\def\arnps#1#2#3#4{\emph{#4}, \emph{ Annu. Rev. Nucl. Part. Sci.} {\bf #1} (#3) #2}
\def\ijmpa#1#2#3#4{\emph{#4}, \emph{ Int. J. Mod. Phys. A} {\bf #1}, #2 (#3)}
\def\jams#1#2#3#4{\emph{#4}, \emph{ J. Austral. Math. Soc. B} {\bf #1} (#3) #2}
\def\appa#1#2#3#4{\emph{#4}, \emph{ Acta Phys. Polonica A} {\bf #1}, (#3) #2}
\def\nat#1#2#3#4{\emph{#4}, \emph{Nature} {\bf #1}, (#3) #2}
\def\science#1#2#3#4{\emph{#4}, \emph{Science} {\bf #1}, (#3) #2}
\def\arcmp#1#2#3#4{\emph{#4}, \emph{Annual Rev. of Cond. Matter Physics} {\bf #1}, (#3) #2}
\def\jcap#1#2#3#4{\emph{#4}, \emph{JCAP} {\bf #1}, (#3) #2}
\def\conphy#1#2#3#4{\emph{#4}, \emph{Contemporary Physics} {\bf #1}, (#3) #2}
\def\ptps#1#2#3#4{\emph{#4}, \emph{ Prog. Theor. Phys. Suppl.} {\bf #1} (#3) #2}
\def\ptp#1#2#3#4{\emph{#4}, \emph{ Prog. Theor. Phys.} {\bf #1} (#3) #2}
\def\apjsup#1#2#3#4{\emph{#4}, \emph{ Astrophys. J. Suppl. Ser.} {\bf #1} (#3) #2}
\def\eurphysjc#1#2#3#4{\emph{#4}, \emph{ Eur. Phys. J.  C} {\bf #1}, #2 (#3)}
\def\njp#1#2#3#4{\emph{#4}, \emph{ New J. Phys. } {\bf #1} (#3) #2}
\def\eurphysjplus#1#2#3#4{\emph{#4}, \emph{ Eur. Phys. J.  Plus} {\bf #1}, #2 (#3)}
%
\def\hepph#1#2{{ hep-ph }{#1} (#2)}
\def\hepth#1#2{{ hep-th }{#1} (#2)}
\def\astroph#1#2{{ astro-ph }{#1} (#2)}
\def\grqc#1#2{{ gr-qc }{#1} (#2)}
\def\ibid#1#2#3#4{\emph{#4}, {\it ibid.} {\bf #1} (#3) #2}

\def\contp#1#2#3#4{\emph{#4}, \emph{ Contemporary Physics} {\bf #1}, #2 (#3)}
\def\physdarkun#1#2#3#4{\emph{#4}, \emph{ Phys. of Dark Universe } {\bf #1}, #2 (#3)}
\def\astrsc#1#2#3#4{\emph{#4}, \emph{Astrophys. Space Sci.} {\bf #1}, #2 (#3)}

\def\epjc#1#2#3#4{\emph{#4}, \emph{ Eur. Phys. J. C} {\bf #1} #2 (#3) }
\def\revphys#1#2#3#4{\emph{#4}, \emph{Reviews in Phys.} {\bf #1} #2 (#3) }

\def\cag#1#2#3#4{\emph{#4}, \emph{ Commun. Anal. Geom.} {\bf #1} #2 (#3) }
\def\contmath#1#2#3#4{\emph{#4}, \emph{ Contemp. Math.} {\bf #1} #2 (#3)}
\def\epjc#1#2#3#4{\emph{#4}, \emph{ Eur. Phys. J. C} {\bf #1} #2 (#3) }
\def\revphys#1#2#3#4{\emph{#4}, \emph{Reviews in Phys.} {\bf #1} #2 (#3) }
\def\science#1#2#3#4{\emph{#4}, \emph{ Science} {\bf #1} #2 (#3) }
\def\pnas#1#2#3#4{\emph{#4}, \emph{ PNAS} {\bf #1} (#3) #2}

\def\jetp#1#2#3#4{\emph{#4}, \emph{J. Exp. Theor. Phys.} {\bf #1} (#3) #2}

\def\pt#1#2#3#4{\emph{#4}, \emph{ Physics Today} {\bf #1}, #2 (#3)}
\def\arnps#1#2#3#4{\emph{#4}, \emph{ Ann. Rev. Nucl. Part. Sci. } {\bf #1} (#3) #2}





\bibitem{han73}
R.S. Hanni and R. Ruffini, \prd{8}{3259}{1973}{Lines of force of a point charge near a Schwarzschild black hole}.
\bibitem{coh71}
 J.M. Cohen and R.M. Wald, \jmp{12}{1845}{1971}{Point charge in the vicinity of a Schwarzschild black hole}.
\bibitem{lin76} 
 B. Linet, \jpm{9}{1081}{1976}{Electrostatics and magnetostatics in the Schwarzschild metric},~
 B. Leaute and B. Linet, \pla{58}{5}{1976}{Electrostatics in a Reissner-Nordstr\"om space-time}.
\bibitem{pet74}
J. A. Petterson, \prd{10}{3166}{1974}{Magnetic field of a current loop around a Schwarzschild black hole}.

 \bibitem{gerruf}
D. Bini, A. Geralico, and R. Ruffini, \pla{360}{515}{2007}{On the equilibrium of a charged massive particle in the field of a Reissner-Nordstr\"om black hole},~
D. Bini, A. Geralico, and R. Ruffini, \prd{75}{044012}{2007}{Charged massive particle at rest in the field of a Reissner-Nordstr\"om black hole}.
\bibitem{wal74}
 R. M. Wald, \prd{10}{1680}{1974}{Black hole in a uniform magnetic field}.
\bibitem{bic80}
J. Bicak and L. Dvorak, \prd{22}{2933}{1980}{
Stationary electromagnetic fields around black holes. 3. General solutions and the fields of current loops near the Reissner-Nordstr\"om black hole}.
\bibitem{cha98}
A. Chamblin, R. Emparan, and G.W. Gibbons, \prd{58}{084009}{1998}{Superconducting p-branes and extremal black holes}.

\bibitem{abd23}
F. Abdulxamidov, J. Rayimbaev, A. Abdujabbarov, and Z. Stuchlik, \prd{108}{044030}{2023}{Spinning magnetized particles orbiting magnetized Schwarzschild black holes} 
\bibitem{qi23}
M. Qi, J. Rayimbaev, and B. Ahmedov, \epjc{83}{730}{2023}{Charged particles and quasiperiodic oscillations around magnetized Schwarzschild black hole}.
\bibitem{ull24}
S. Ullah Khan, O. Abdurkhmonov, J. Rayimbaev, S. Ahmedov, Y. Turaev, and S. Muminov, \epjc{84}{650}{2024}{Circular motion and collisions of particles with magnetic dipole moment
and electric charge in dipolar magnetosphere around Schwarzschild black holes}.

\bibitem{fro10}
V. P. Frolov and A. A. Shoom, \prd{82}{084034}{2010}{Motion of charged particles near a weakly magnetized Schwarzschild black hole}.
\bibitem{fro12}
V. P. Frolov, \prd{85}{024020}{2012}{Weakly magnetized black holes as particle accelerators}. 

\bibitem{zah13}
A. M. Al Zahrani, V. P. Frolov, and A. A. Shoom, \prd{87}{084043}{2013}{Critical escape velocity for a charged particle moving around a weakly magnetized Schwarzschild black hole}.

\bibitem{fel03}
F. de Felice and F. Sorge, \cqg{20}{469}{2003}{Magnetized orbits around a Schwarzschild black hole}.

\bibitem{ray16}
J. R. Rayimbaev, \astrsc{361}{288}{2016}{Magnetized particle motion around non-Schwarzschild black hole immersed in an external uniform magnetic field}.

\bibitem{abd20}
A. Abdujabbarov, J. Rayimbaev, B. Turimov, and F. Atamurotov, \physdarkun{30}{100715}{2020}{Dynamics of magnetized particles around 4-D Einstein Gauss-Bonnet black hole}. 

\bibitem{rog25a}
M. Rogatko and P. Verma, \prd{111}{084042}{2025}{Influence of dark photon on magnetized and charged particle orbits around static spherically symmetric black hole}.
\bibitem{rog25b}
M. Rogatko and P. Verma, \epjc{85}{205}{2025}{Can synchrotron radiation reveal the presence of dark sector around black hole?}.







\bibitem{gar91}
D. Garfinkle and S-J. Roy, \plb{257}{158}{1991}{Angular momentum of an electric charge and magnetically charged black hole}.
\bibitem{kim07}
J. H. Kim and S-H. Moon, \jhep{09}{2007}{088}{Electric charge in interaction with magnetically charged black holes}.
\bibitem{bun07}
C. Bunster and M. Henneaux, \pnas{104}{12243}{2007}{A monopole near a black hole}.


\bibitem{bru09}
F. Br\"ummer and J. Jaeckel, \plb{675}{360}{2009}{Minicharges and magnetic monopoles}.





\bibitem{fil23}
M. Filzinger, S. D\"orscher, R. Lange, J. Klose, M. Steinel, E. Benkler, E. Peik, C. Lisdat, and N. Huntemann, \prl{130}{253001}{2023}
{Improved Limits on the Coupling of Ultralight Bosonic Dark Matter to Photons from Optical Atomic Clock Comparisons}.
\bibitem{ram23}
K. Ramanathan, N. Klimovich, R. Basu Thakur, B.H. Eom, H.G. Leduc, S. Shu, A.D. Beyer, and P.K. Day,
\prl{130}{231001}{2023}{Wideband Direct Detection Constraints on Hidden Photon Dark Matter with the QUALIPHIDE Experiment}.
\bibitem{kot23}
 S. Kotaka, S. Adachi, R. Fujinaka, S. Honda, H. Nakata, Y. Seino, Y. Sueno, T. Sumida, J. Suzuki, O. Tajima, and S. Takeichi, \prl{130}{071805}
 {2023}{Search for Dark Photon Dark Matter in the Mass Range $74-110\text{ }\text{ }\mathrm{\ensuremath{\mu}}\mathrm{eV}$ with a Cryogenic Millimeter-Wave Receiver}.






\bibitem{hol86}
B. Holdom, \plb{166}{196}{1986}{Two $U(1)$'s and $\ep$ charge shifts}.
\bibitem{cap21}
A. Caputo, A.J. Miller, C.A.J. O'Hare, and E. Vitagliano, \prd{104}{095029}{2021}{Dark photon limits: A handbook}.




\bibitem{ach16}
B.S. Acharya, S.A.R. Ellis, G.L. Kane, B.D. Nelson, and M.J. Perry, \prl{117}{181802}{2016}{Lightest Visible-Sector Supersymmetric Particle is Likely Unstable}.





\bibitem{jea03}
P. Jean {\it et al.}, \aa{407}{L55}{2003}{Early SPI/INTEGRAL measurements of 511 keV line emission from the 4th quadrant of the Galaxy}.
\bibitem{cha08}
J. Chang {\it et al.}, \nat{456}{362}{2008}{An excess of cosmic ray electrons at energies of 300-800 GeV}.
\bibitem{bub14}
E. Bulbul et al., \apj{789}{13}{2014}{Detection of an unidentified emission line in the stacked X-ray spectrum of galaxy clusters}.

\bibitem{fil20}
A. Filippi and M. De Napoli, \revphys{5}{100042}{2020}{Searching in the dark: the hunt for the dark photon}.
\bibitem{ger15}
A. Geringer-Sameth and M.G. Walker, \prl{115}{081101}{2015}{Indication of Gamma-Ray Emission from the Newly Discovered Dwarf Galaxy Reticulum II}.
\bibitem{bod15}
K.K. Boddy and J. Kumar, \prd{92}{023533}{2015}{Indirect detection of {\it dark matter} using MeV-range gamma-rays telescopes}.
\bibitem{til15}
K.Van Tilburg, N. Leefer, L. Bougas, and D. Budker, \prl{115}{011802}{2015}{Search for Ultralight Scalar Dark Matter with Atomic Spectroscopy}.
\bibitem{cha17}
J.H. Chang, R. Essig, and S.D. McDermott, \jhep{01}{2017}{107}{Revisiting Supernova 1987A constraints on dark photons}.
\bibitem{sensei}
M. Crisler et. al. (SENSEI Collaboration), \prl{121}{061803}{2019}{SENSEI: First Direct-Detection Constraints on Sub-GeV Dark Matter from a Surface Run}
\bibitem{lee14}
J.P. Lees et al., \prl{113}{201801}{2014}{Search for a Dark Photon in $e^+ e^-$ Collisions at BABAR}.
\bibitem{dav11}
M. Davier et al., \epjc{71}{1515}{2011}{Reevaluation of the hadronic contributions to the muon g-2 and to $\alpha(M^2_z)$}. 

\bibitem{alo25}
G. Alonso-Alvarez and D. Curtin, \jcap{05}{082}{2025}{Dark astronomy with dark matter detectors}.







\bibitem{ber18b}
G. Bertone and T.M.P. Tait, \nat{562}{51}{2018}{A new era in the search for dark matter}.



\bibitem{kic19}
B. Kiczek and M. Rogatko, \jcap{09}{049}{2019}{Ultra-compact spherically symmetric dark matter charged star}.
\bibitem{kic20}
B. Kiczek and M. Rogatko, \prd{101}{084035}{2020}{Influence of black matter on black scalar hair}.
\bibitem{kic21}
B. Kiczek and M. Rogatko, \prd{103}{124021}{2021}{Axion-like dark matter clouds around rotating black holes}.
\bibitem{kic22}
B. Kiczek and M. Rogatko, \epjc{82}{586}{2022}{Static axion-like dark matter clouds around magnetized rotating wormholes - probe limit case}.



\bibitem{dark mon}
W. Fischler and W. Tangarife Garcia, \jhep{01}{2011}{025}{Hierarchies of SUSY Splittings and Invisible Photinos as Dark Matter},~
C. Gomez Sanchez and B. Holdom, \prd{83}{123524}{2011}{Monopoles, strings and dark matter},~
S. Baek, P. Ko and W.-I. Park, \jcap{10}{067}{2014}{Hidden sector monopole, vector dark matter and dark radiation with Higgs portal},~
V.V. Khoze and G. Ro, \jhep{10}{2014}{061}{Dark matter monopoles, vectors and photons},~
V.V. Burdyuzha, \jetp{127}{638}{2018}{Magnetic Monopoles and Dark Matter}.





\bibitem{ter19}
J. Terring and C. B. Verhaasen, \jhep{12}{2019}{152}{Detecting dark matter with Aharonov-Bohm}.

\bibitem{gra22}
M. L. Graerrer, I. M. Shoemaker, and N. T. Arellano, \jhep{03}{2022}{105}{Milli-magnetic monopole dark matter and the survival of galactic magnetic fields}.
\bibitem{ven21}
V. Vento, \epjc{81}{2021}{229}{Primordial monopolium as dark matter}.
\bibitem{yan23}
J. Yang, R. Zhou, and L. Bian, \plb{839}{137822}{2023}{Gravitational waves and monopoles from first-order phase transitions}.

\bibitem{hir21}
T. Hiramatsu, M. Ibe, M. Suzuki, and S. Yamaguchi, \jhep{12}{2021}{122}{Gauge kinetic mixing and dark topological defects}.
\bibitem{chi23}
A. Chitose and M. Ibe, \prd{108}{035044}{2023}{Interactions of electrical and magnetic charges and dark topological defects}.


\bibitem{pre84}
J. Preskill, \arnps{34}{461}{1984}{Magnetic monopoles}.
\bibitem{aha59}
Y. Aharonov and D. Bohm, \pr{115}{485}{1959}{Significance of electromagnetic potentials in the quantum theory}.

\bibitem{fel}
B. Felsanger, {\it Geometry, Particles, and Fields}, Springer, New York 1998.

\bibitem{abexp}
R.G. Chambers, \prl{5}{3}{1960}
{Shift of an Electron Interference Pattern by Enclosed Magnetic Flux},~
A. Tonomura {\it et al.}, \prl{56}{792}{1986}{Evidence for Aharonov-Bohm effect with magnetic field completely shielded from electron wave}.

\bibitem{raj16}
A. Rajantie, \pt{69}{41}{2016}{The search for magnetic monopoles}.

\bibitem{ann09}
J.F. Annett, {\it Superconductivity, Superfluids and Condensates}, Oxford University Press, Oxford 2009.





\bibitem{rog24}
M. Rogatko, \prd{109}{104030}{2024}{Dark photon-dark energy stationary axisymmetric black holes}.


\bibitem{gia03}
G. Giacomelli and L. Patrizii, {\it Magnetic monopole searches}, hep-ex 0302011 (2003).


\end{thebibliography}
\end{document}